\documentclass[prd,twocolumn,nofootinbib]{revtex4}

\usepackage{amstext,amsmath,amssymb}
\usepackage[dvips]{graphicx}
\usepackage{latexsym}

\newcommand{\Ref}[1]{(\ref{#1})}
\def \subs{\subsection}

\newcommand{\Z}{\mathbb{Z}}

\newcommand{\R}{\mathbb{R}}

\def\be{\begin{equation}}
\def\ee{\end{equation}}
\def\bes{\begin{eqnarray}}
\def\ees{\end{eqnarray}}
\def\nn{\nonumber}
\def\arr{\rightarrow}

\def\om{\omega}

\def\la{\langle}
\def\ra{\rangle}
\def\f{\frac}
\def\tl{\tilde}

\def\what{\widehat}

\def \vphi{\varphi}

\def\pp{\partial}
\def\hh{{\cal H}}
\def\mn{{\mu\nu}}
\def\ii{{\cal I}}
\def\ff{{\cal F}}
\def\cc{{\cal C}}
\def\dd{{\cal D}}
\def\oo{{\cal O}}

\def\eps{\epsilon}
\def\ka{\kappa}

\newcommand{\lalg}[1]{\mathfrak{#1}}

\newcommand{\SO}{\mathrm{SO}}

\renewcommand{\sl}{\lalg{sl}}

\newcommand{\so}{\lalg{so}}

\def\dag{^\dagger}
\def\eps{\epsilon}

\def\xx{{\cal X}}

\newcommand{\ket}[1]{|#1\rangle}

\begin{document}

\title{The Relativistic Particle: Dirac observables and Feynman propagator}

\author{{\bf Laurent Freidel}\footnote{lfreidel@perimeterinstitute.ca}}
\affiliation{Perimeter Institute, 31 Caroline St N., Waterloo, ON, Canada N2L 2Y5}
\author{{\bf Florian Girelli}\footnote{girelli@sissa.it}}
\affiliation{SISSA, Via Beirut 2-4, 34014 Trieste, Italy and INFN, Sezione di Trieste}
\author{{\bf Etera R. Livine}\footnote{etera.livine@ens-lyon.fr}}
\affiliation{Laboratoire de Physique, ENS Lyon, CNRS UMR 5672, 46 All\'ee d'Italie, 69364 Lyon Cedex 07}

\begin{abstract}

We analyze the algebra of Dirac observables of the relativistic particle in four
space-time dimensions. We show that the position observables become non-commutative and
the commutation relations lead to a structure very similar to the non-commutative
geometry of Deformed Special Relativity (DSR). In this framework, it appears natural to
consider the 4d relativistic particle as a five dimensional massless particle. We study
its quantization in terms of wave functions on the 5d light cone. We introduce the
corresponding five-dimensional action principle and analyze how it reproduces the physics
of the 4d relativistic particle. The formalism is naturally subject to divergences and we
show that DSR arises as a natural regularization: the 5d light cone is regularized as the
de Sitter space. We interpret the fifth coordinate as the particle's proper time while
the fifth moment can be understood as the mass. Finally, we show how to formulate the
Feynman propagator and the Feynman amplitudes of quantum field theory in this context in
terms of Dirac observables. This provides new insights for the construction of
observables and  scattering amplitudes in DSR.
\end{abstract}

\maketitle



\section*{Introduction}


There has recently been an increasing interest in theories of Deformed Special Relativity
(DSR). Tentatively introduced as Lorentz invariant theories with modified dispersion
relation taking into account a universal length (or mass) scale \cite{DSR}, they are
believed to provide grounds for a phenomenology of quantum gravity in the
semi-classical regime. In three space-time dimensions, it has been shown that matter
degrees of freedom are described after integration over the metric fluctuations by an
effective non-commutative quantum field theory which provides an explicit realization of
a DSR theory \cite{eteralaurent}. There also are several heuristic arguments for 4d
quantum gravity \cite{DSR4d}, even though we do not yet have a definitive derivation of a
DSR quantum field theory from quantum gravity. It is nevertheless important to understand
how to build a consistent quantum field theory based on such Deformed Special Relativity.
On one hand, it would provide us  modifications of scattering amplitudes which could be
tested experimentally in particle accelerators or in cosmological context; on the other
hand, it might provide us some insights in the structure of a full quantum gravity theory.

DSR is usually presented as a theory based on a curved momentum space: the momentum does
not live in the standard flat 4d Minkowski space but in the de Sitter space (e.g.
\cite{jerzy}). This curvature induces by duality the non-commutativity of the space-time
coordinates. The goal is to write a quantum field theory on such a background. Our
strategy is to re-examine Special Relativity (SR) and the structure of the algebra of
observables of the relativistic particle in order to understand its extension to DSR and
provide the deformed theory with solid foundations.

Starting with a standard massive relativistic particle, we first construct the set of
strong Dirac observables. These are the phase space functions which commute everywhere in phase space with the Hamiltonian
constraint.  These ``constants of motion" correspond to the  measurable quantities. In this
simple case, they are generated by the 4-momentum $p_\mu$ and the Lorentz generators
$j_\mn$. Together, they generate the Poincar\'e Lie algebra. However, if we are
interested in probing the structure of space-time, we would like to identify suitable
space-time coordinates which are Dirac observables. The coordinates $x_\mu$ obviously are
not observables. To construct good coordinate functions, we use the concept of relational
observables: we choose one of the degree of freedom of the system as the clock and we
describe the evolution of the remaining degrees of freedom in term of that internal time.
One usually chooses the time coordinate $x_0$ as the clock. This leads to the
Newton-Wigner position operators \cite{newton}. They are Dirac observables, but they are not Lorentz
covariant. Thus we chose to work with the Lorentz invariant clock $x_\mu p^\mu$. They
lead to well-defined Lorentz covariant position observables $\xx_\mu$. Together with the
$p^\mu$'s, these new coordinates generate the whole algebra of observables.

Observing that using   $\xx_\mu$ leads to the impossibility to define a time evolution as well as complications  in the quantization procedure, we extend the analysis to the Lorentz covariant position weak observables $X_\mu$, that is the position observables that commute with the Hamiltonian constraint only on shell. 

In section II, we recall how the choice of scalar product is of fundamental importance to define the quantum observables: dealing with the kinematical or the physical scalar product leads to different result for the self adjoint position operator $\what X_\mu$.

In section III, we recall  how the  $\what X_\mu$ non-commutativity  reflects
the impossibility of localizing the quantum relativistic particle with an accuracy better than
the Compton length. This non-commutativity turns out to be very similar to the one
encountered in DSR. This initial observation shows that Special Relativity already
contains the seeds of its extension to DSR.

In section IV, we show that the algebra of the $X,p,j$ observables is naturally
quantized as operators acting on the space of functions on the five-dimensional light
cone: the massive 4d relativistic particle becomes a massless 5d system. We also make
explicit the isomorphism between this new space of wave functions on the 5d light cone
and the standard wave functions on the flat Minkowski space.

In section V, we further introduce a 5d action principle for the 4d massive relativistic
particle. We study the map between the new 5d coordinates and the usual 4d coordinates
$(x,p)$.  It appears that the fifth moment actually generates the Hamiltonian flow of the
relativistic particle: this new fifth component of the momentum can be considered as the
(rest) mass of the particle.

However, most of the Poisson (and Dirac) brackets induced in 4d become singular on the 5d
light cone. Therefore, we introduce a regularization slightly moving away from the 5d
light cone: the 5d momentum now lives on the de Sitter space. The resulting modified 5d
action has been shown in \cite{GKKL} to generate the DSR theories as different gauge
fixing choices. From this point of view, we here show that DSR appears as a natural
regularization of SR at the level of the algebra of observables.

In section VI, we exploit the 5d reformulation of the relativistic particle to write a 5d
representation of the Feynman propagator (for a massive scalar field). Once again, the
expression becomes singular on the 5d light cone. Nevertheless, we show that the Feynman
propagator can be written exactly as a integral on the de Sitter space of moments. This
analysis leads to interpreting the fifth space-time coordinate as the proper time of the
particle.

Going further in the analysis of the Quantum Field Theory amplitudes, we show  how the
Feynman loop diagram evaluations can be written as expectation values of some (time
ordered) Dirac observables of the relativistic particle.  We hope to be able to
generalize this to DSR. This would be a definite first step towards deriving the Feynman
amplitudes in DSR and constructing a consistent S-matrix describing the scattering of
particles.

In the last section, we introduce the necessary framework to deal with particles with
spin. We perform the canonical analysis and write the corresponding Dirac observables. We
insist on the fact that the spin already induces a (Moyal-like) non-commutativity of the
space-time coordinates at the classical level. This non-commutativity is distinct from
the non-commutativity of the position observables (which is of a $\kappa$-deformed
Poincar\'e type) and leads to further difficulties in the quantization of the algebra of
Dirac observables.

\section{Dirac Observables for the Relativistic Particle}

\subsection{Position observables}\label{IA}

We start with the phase space of the relativistic particle: \be
\{x_\mu,p_\nu\}=\eta_{\mu\nu}, \ee where we choose the flat metric $\eta_\mn=(+---)$. The
Hamiltonian constraint for a massive particle is $H=(p^2-m^2)\equiv 0$ and the action is:
\be S=\int p^\mu dx_\mu -\lambda H, \ee where $\lambda$ is a Lagrange multiplier. We are
interested into the Dirac observables,  which are the phase space functions which
Poisson-commute with $H$. They are generated by the momenta $p_\mu$, which generate the
translations, and the generators of the Lorentz transformation: \be j_{\mu\nu}=x_\mu
p_\nu-x_\nu p_\mu. \ee Nevertheless, we would like to have some Dirac observables giving
the position of the particle. For this purpose, we use relational observables
\cite{carlo}. These are constructed from two arbitrary phase space functions $a,b$ and
defined as a function of an arbitrary real parameter $T$: \be A_b(T)\,\equiv\, \int_\R
d\tau
\,a(\tau)\,\dot{b}(\tau)\delta(b(\tau)-T). \ee We have introduced the notation $f(\tau)$ for the Hamiltonian flow of
the phase space function $f$:
$$
f(\tau)\,\equiv\, e^{-\f12\tau\{H,\cdot\}}\,f.
$$
It is straightforward to check that $A_b(T)$ is a Dirac observable, whatever the value of
$T$.  It represents the value of $a$ when the clock $b$ indicates $T$. It is then natural
to choose a clock $b=x\cdot v$ with $v$ an arbitrary fixed (time-like) vector and the
function $a=x_\mu$ indicating the particle's position. This way, we define the following
Dirac observables: \be \xx_\mu^{(v)}(T) =x_\mu+\f{p_\mu}{p\cdot v}\left(T-x\cdot v\right)
=\f{j_{\mu\nu}v^\nu+Tp_\mu}{p\cdot v}, \ee which defines the values of the four coordinates at the time $x^\alpha
v_\alpha=T$. When $v=(1,0,0,0)$, the time is simply the time coordinate $x_0$. These position observables commute with
each other:
$$
\{\xx_\mu^{(v)}(T),\xx_\nu^{(v)}(T)\}=0.
$$

However, the $\xx_\mu^{(v)}(T)$ are not Lorentz-covariant, since they are constructed
using a fixed vector $v_\mu$. The simplest solution to address this problem is to take
the special vector $v\equiv p$: we choose as clock the dilatation $b=x\cdot p = D$. This
leads to the following position observables:
\be
\xx_\mu(T)=x_\mu+\f{p_\mu}{p^2}(T-D)=\f{k_\mu+T p_\mu}{p^2},
\label{Xobs}
\ee with
$$
k_\mu=j_{\mu\nu}p^\nu=x_\mu p^2 -p_\mu D.
$$
To understand the physical meaning of these coordinates, it is useful to write them as:
\be
\xx_\mu(T)=\left(\eta_{\mu\nu}-\f{p_\mu p_\nu}{p^2}\right)x^\nu+ T \f{p_\mu}{p^2}.
\ee
We always have $\xx(T)\cdot p=T$ and the coordinates $\xx_\mu(T=0)$ are simply the
transversal coordinates of the vector $x$ wit respect to the particle trajectory. More
precisely $D=x\cdot p=0$ corresponds to the 'perihelion' $P$ of the particle's
trajectory, i.e. the event when the particle is the closest to the origin. Then $T$
counts the proper time along the particle's trajectory from $P$.

Let us work at some fixed time $T_0$. The coordinates $x_\mu$ can not be reconstructed
solely from the variables $\xx_\mu(T_0),p_\mu$ since they are Dirac observables and
$x_\mu$ is not. To invert the relation between the $x$'s and the $\xx(T_0)$'s, we need
the dilatation $D$, which is not a Dirac observable\footnotemark. Then specifying the
four coordinates $x_\mu$ is equivalent to specifying the 5-vector $(D,\xx_\mu(T_0))$,
which is the parallel and transversal  projections of $x$ with respect to the vector $p$
(up to the shift $T_0p_\mu$).

\footnotetext{Indeed $\{D,p^2\}$ does not vanish. More precisely, one can check:
$$
\{D,p^2\}=2p^2,\quad \{D,x^2\}=-2x^2,\quad \{p^2,x^2\}=-4D,
$$
so that $D/2$, $p^2/2$ and $x^2/2$ form a $\sl(2,\R)$ algebra.
This is the starting point of 2-time physics \cite{2time}. }

Let us look at the algebra generated by the $\xx_\mu(T)$.
The key remark is that the $j_{\mu\nu}$'s and $k_\mu$'s form a $\so(1,4)$ algebra:
\be
\{k_\mu,k_\nu\}=-(p^2) j_{\mu\nu}.
\ee
This implies that the $\xx_\mu$'s do not commute with each other:
\be
\{\xx_\mu(T),\xx_\nu(T)\}=
-\f{j_{\mu\nu}}{p^2}.
\ee
We also have a deformation of the canonical Poisson bracket:
\be
\{\xx_\mu(T),p_\nu\}=\eta_{\mu\nu}-\f{p_\mu p_\nu}{p^2}.
\ee
This is the projector onto transversal modes (orthogonal to the direction of the motion $p_\mu$).
Finally, we compute the action of the dilatation:
$$
\{D,\xx_\mu\}=-\xx_\mu,\quad
\{D,p_\mu\}=+p_\mu,\quad
\{D,j_\mn\}=0.
$$
The essential point for our discussion is that the $j_\mn$ and the rescaled\footnote{Since the
$\xx$'s commute with the Hamiltonian, we can rescale them by any function of $p^2$ without
complicating their Poisson brackets. Moreover the coordinate choice $\sqrt{p^2}\xx_\mu$ is
invariant under dilatations generated by $D$: these are the natural coordinates to consider when
using $D$ as time \cite{GKKL}.}  positions $\sqrt{p^2}\xx_\mu$ form a $\so(4,1)$ algebra under the
Poisson bracket. This is reminiscent of the Snyder algebra for a Lorentz covariant non-commutative
geometry \cite{snyder}. Actually this link can be made precise as we will see in the next section
\ref{weakobs}. The interesting point is that the Snyder algebra is related (through a change of
basis) to the $\kappa$-deformed Poincar\'e algebra encountered in theories of Deformed Special
Relativity \cite{GKKL}. In some sense, we show here that standard Special Relativity itself already
contains the seeds of DSR.

We now compute the values of the two Casimir operators of the $\so(4,1)$ algebra. For the quadratic Casimir, we find\footnote{We
compute the Lorentz invariant:
$$
\xx_\mu \xx^\mu\,=\, x_\mu x^\mu +\f{1}{p^2}(T^2-D^2).
$$}:
\be
{\cal C}_2\equiv
\f{1}{2}j_\mn j^\mn -p^2 \xx_\mu(T) \xx^\mu(T) =-T^2.
\ee
The time becomes a Casimir of our algebra of Dirac observables $(j,\xx(T))$.
Then, introducing the Pauli-Lubanski vector $\om_\mu\equiv \epsilon_{\mu\alpha\beta\gamma}\xx^\alpha j^{\beta\gamma}=0$, the
quartic Casimir turns out to be trivial: \be {\cal C}_4\equiv\om_\mu\om^\mu=0. \ee This means that we are dealing with a simple
representation of the algebra $\so(4,1)$, which can be realized as functions on the 5d light cone $\cc_0$, on the one-sheet
hyperboloid $\SO(4,1)/\SO(3,1)$ (which actually is the de Sitter space) or on the two-sheets hyperboloid $\SO(4,1)/\SO(4)$.

At the level of the action, we compute the kinetic term $p^\mu dx_\mu$ in term of the
coordinates $\xx_\mu(T)$ and find:
\be
p^\mu dx_\mu= p^\mu d\xx_\mu + \f{(T-D)}{2p^2}dH+dD.
\ee
Therefore the action can be written up to total derivatives as:
\be
S'=\int p^\mu d\xx_\mu + Hd\left(\f{D-T}{2p^2}\right)-\lambda H-\mu (p^\mu \xx_\mu -T).
\ee
This leads to two considerations. First, $D/p^2$ can be considered
as a new fifth coordinate, conjugate to the Hamiltonian constraint
$H=p^2-m^2$. It thus seems possible to provide the relativistic
particle with a five-dimensional action principle. Second the five
dimensions are reduced to the usual four dimensions by an extra
constraint. This second constraint do not commute with the
Hamiltonian constraint $H=0$ and can therefore be considered as a
gauge fixing condition lifting the first class constraint $H$ to a
second class constraint system. This is analyzed in details in section \ref{5dquant}.

\subsection{Strong observables versus weak observables}
\label{weakobs}

Up to now, we have considered {\it strong observables}, that commute exactly with the
Hamiltonian constraint. We would like to propose to use {\it weak observables} instead,
that commute with the Hamiltonian constraint only on the mass-shell.

Indeed, first, from the point of view of the quantization, the factors $1/p^2$ and
$\sqrt{p^2}$ occurring when considering the Dirac observables $\xx_\mu(T)$ would not be
easy to deal with when quantizing the algebra of observables. Then, from the point of the
dynamics, the "time" evolution in $T$ of the coordinates $\xx_\mu(T)$ can not be
generated from a Hamiltonian. More precisely, there does not exist any phase function
$H_{eff}$ such that:
$$
\forall T,\,\{H_{eff},\xx_\mu(T)\}=\f{d\xx_\mu(T)}{dT}=\f{p_\mu}{p^2}.
$$
This is due to the fact that the coordinates $\xx_\mu(T)$ are basically the projection of
the space-time coordinates $x_\mu$ orthogonally to the momentum $p_\mu$. We can relax
this condition by introducing the following weak observables:
$$
X_\mu(T)=\left(\eta_\mn-\f{p_\mu p_\nu}{m^2}\right)x^\nu +T\f{p_\mu}{m^2}
=x_\mu + \f{p_\mu}{m^2}(T-x^\nu p_\nu).
$$
In contrast to the previous situation, this relation is invertible off shell\footnote{This is due to the fact that $p.X$ is not a constant anymore but is easily related to $p.x$:
$$
p_\nu X^\nu =p_\nu x^\nu \left(1-\f{p^2}{m^2}\right).
$$} and we
can express $x_\mu$ in terms of $X_\mu(T=0)$:
\be
x_\mu\,=\, \left(\eta_\mn-\f{p_\mu p_\nu}{p^2-m^2}\right)X_\mu.
\ee

These do not strongly commute with the Hamiltonian constraint anymore:
\be
\{H,X_\mu(T)\}\,=\,
-2p_\mu\left(1-\f{p^2}{m^2}\right)\,=\, 2\f{p_\mu}{m^2}H,
\ee
but the Poisson bracket still vanishes on-shell. The $X_\mu(T)$ still do not commute with
each other:
\be
\{X_\mu(T),X_\nu(T)\}\,=\,-\f{j_\mn}{m^2}.
\ee
The $(p,X,j)$ algebra  is very similar to the previous $(p,\xx,j)$ algebra, but the $1/p^2$ factors
are replaced by constant $1/m^2$ factors. The observables $j_\mn$ and $mX_\mu$ form a $\so(4,1)$
algebra. Moreover, the $(p,X,j)$ algebra is now exactly the Snyder algebra related to a
$\kappa$-deformed Poincar\'e symmetry with deformation parameter $\kappa\equiv m$. This provides an
exact link between the algebra of Dirac observables of the relativistic particle and the phase
space structure of the deformed relativistic particle.

A further advantage of the $X$ observables on the $\xx$ coordinates is that the $p^2$ factors are
replaced by $m^2$ constants: we do not have to deal with any $\sqrt{p^2}$ factor when quantizing
(compare $\sqrt{p^2}\xx_\mu$ to the simpler $mX_\mu$).
Of course, other subtleties will arise. First, although the quartic Casimir will still vanish, the
quadratic Casimir $C_2$ will only be equal to $-T^2$ on the mass-shell. Then, as we will discuss in
the paragraphs below, although the operators $\what{X_\mu}$ will of course be Hermitian for the
physical scalar product, they will not be Hermitian with respect to the kinematical scalar product.

Moreover, it is possible to generate the time evolution of the observables $X_\mu(T)$
with an effective Hamiltonian which turns out to be exactly the logarithm of the original
Hamiltonian constraint:
\be
\left\{\,\f12\ln|H|,X_\mu(T)\right\} =\f{p_\mu}{m^2} =\f{dX_\mu(T)}{dT}.
\label{Heff}
\ee

Finally, we can re-write the action in term of these new coordinates (up to a total derivative):
\be
p^\mu dx_\mu\,=\, p^\mu dX_\mu -\f12 (D-T)dH,
\ee
which shows explicitly the canonical relation between the Hamiltonian constraint and the
dilatation generator.

\subsection{Gauge fixing and the Dirac bracket}

The choice of a time variable can be understood as an explicit gauge fixing that breaks
the symmetry of the action under time reparametrization. After gauge fixing, the
symplectic form on the reduced phase space is given by the Dirac bracket. Given the
constraint $H$ and a gauge fixing condition $C$ such that $\{C,H\}\neq0$, the Dirac
bracket is defined as
\begin{eqnarray}
\{\phi,\psi\}_D & = & \{\phi,\psi\}
-\{\phi,C\}\left(\frac{1}{\{H,C\}}\right)\{H,\psi\} \nn\\
&& -\{\phi,H\}\left(\f{-1}{\{H,C\}}\right)\{C,\psi\}.
\end{eqnarray}

For the standard time choice $\cc=x.v$, where $v$ is an arbitrary time-like vector, we
have $\{H,x.v\}=-2p.v$ and we obtain:
\begin{eqnarray}
\{x_\mu,p.v\}_D&=&\left(\eta_\mn-\f{p_\mu v_\nu}{p.v}\right)v^\nu \nn\\
\{x_\mu,x_\nu\}_D&=&0.
\end{eqnarray}
$p.v$ acts as the Hamiltonian on the gauge fixed system: the time $x.v$ commute with
$p.v$ and the space coordinates (orthogonal to $v$) evolve with the usual speed defined
by the momentum $p_\mu$. The important relation linking the relational Dirac observables
written in the previous section and the gauge fixing procedure is the equality between
the Dirac bracket and the Poisson bracket of the observables:
\begin{eqnarray}
\{x_\mu,p_\nu\}_D&=&\{\xx^{(v)}_\mu(T),p_\nu\}, \nn\\
\{x_\mu,x_\nu\}_D&=&\{\xx^{(v)}_\mu(T),\xx^{(v)}_\nu(T)\},
\end{eqnarray}
for any value of the parameter $T$.


Choosing the dilatation as gauge fixing condition, $C=D$, we compute
\begin{eqnarray}
\{x_{\mu},p_{\nu}\}_D&=&\eta_{\mu\nu}- \frac{p_{\mu}p_{\nu}}{p^2}, \nn\\
\{x_{\mu},x_{\nu}\}_D&=&-\frac{j_{\mu\nu}}{p^2}.
\end{eqnarray}
And we similarly obtain the following equalities:
\begin{eqnarray}
\{x_\mu,p_\nu\}_D&=&\{\xx_\mu(T),p_\nu\}, \nn\\
\{x_\mu,x_\nu\}_D&=&\{\xx_\mu(T),\xx_\nu(T)\}.
\end{eqnarray}

\section{Quantization of the Weak Observables: hermiticity}

We are interested in the Hermicity properties of the quantum operator $\what X_\mu$. As
we mentioned earlier, since $X_\mu(T)$ are only weak observables, their hermiticity
properties will differ depending if we consider the kinematical inner product or the
physical inner product.

In the following, we will focus on $X_\mu(T=0)$ which we will simply denote $X_\mu$. The
other term $+Tp_\mu/m^2$ can be easily taken into account. We work in the
$p$-polarisation: wave functions are functions of the momentum $p_\mu$, $\what p$ acts
by multiplication while $\what x_\mu=i\pp/\pp p$ acts as a derivation operator. The
kinematical inner product is simply defined as $\la \psi|\phi \ra\,=\,\int d^4p\,
\overline{\psi}(p)\phi(p)$, while the physical inner product takes the Hamiltonian constraint into
account:
$$
\la \psi|\phi \ra_{ph}\,=\,
\int d^4p_\mu \,\delta(p^2-m^2)\,\overline{\psi}(p)\phi(p).
$$
Then, if we choose the trivial ordering for $\what{X_\mu}$:
\be
\what X_\mu\,=\, i\,\f{\pp}{\pp p^\mu}-\,i\,\f{p_\mu}{m^2}p_\nu\f{\pp}{\pp p_\nu},
\ee
a straightforward calculation gives:
$$
(\what X_\mu\dag)_{kin}\,=\,\what X_\mu-\f{i(d+1)}{m^2}p_\mu,
$$
\be
(\what X_\mu\dag)_{ph}\,=\,\what X_\mu-\f{i(d-1)}{m^2}p_\mu,
\ee
where $d=4$ is the space-time dimension. Therefore, at the quantum level, for the
physical inner product, the correct {\it self-adjoint} operator representing the
observable $X_\mu(T)$ requires a complex shift:
\be
\what X_\mu(T)\,\equiv\, i\,\pp_\mu-\,i\,\f{p_\mu p_\nu}{m^2}\pp^\nu
+\left(T-i\f{(d-1)}{2}\right)\f{p_\mu}{m^2},
\label{timeshift}
\ee
where $\pp_{\alpha}$ is the partial derivative with respect to the momentum variable
$p_{\alpha}$.

In the following section, we will quantize the relativistic particle using the $\so(4,1)$
structure of the algebra of observables and we will check that we recover the exact same
shift from the 5d perspective.


\section{On the Localization of the quantum particle}

Since we know the quantization of the position observables
$\what{X}_\mu$,  we can study the issue of the localization of the
relativistic particle at the quantum level in terms of Dirac
observables.

$\what{X}_\mu(T)$ is quantized in term of the $\so(4,1)$ generator $J_{\mu 4}/m$. In the space-like sector $\mu=i=1,2,3$, the
spectrum of $\what{X}_i$ is discrete, $\hbar/m\,\Z$, and space distances are quantized in units of Compton length $l_C=\hbar/m$.
On the other hand,  the spectrum of the time coordinate $\what{X}_0\,\propto J_{0 4}$ remains continuous.

Furthermore, one can compute the exact spectrum of the distance
operator $\what X_\mu \what X^\mu$ \cite{dan1} and one shows that the negative
eigenvalues (corresponding to the space-like sector) are discrete
while its positive eigenvalues are continuous (corresponding to
the time-like sector). As explained in \cite{dan1}, the
discreteness of the distances does not contradict the Lorentz
invariance of the theory.

Let us point out that $X_\mu X^\mu$ is not the usual metric $x_\mu x^\mu$. Nevertheless, it is a Lorentz invariant Dirac
observable, which coincides with $x_\mu x^\mu$ when $D$ is fixed i.e when working in  fixed eigenspace of the dilatation
$\what{\dd}$. Indeed we recall that:
$$
X_\mu(T) X^\mu(T)\,=\,x_\mu x^\mu +\f{1}{p^2}(T^2-D^2).
$$
This discrete lattice-like structure of the coordinates $X_\mu$ naturally leads to some intrinsic uncertainties in the
measurement of these position Dirac observables. Indeed, as shown in section \ref{IA}, the commutator of the coordinates reads:
\be \{X_\mu,X_\nu\}= -\f{1}{\hbar}(l_C)^2\f{j_{\mu\nu}}{\hbar}, \ee where $l_C=\hbar/m$ is the Compton length of the particle (at
rest). This Poisson bracket will get quantized as:
$$
[\what{X}_\mu,\what{X}_\nu]=i\,(l_C)^2 \,\what{\jmath}_{\mu\nu}.
$$
From this, we expect a position uncertainty $\delta X \sim l_C$,
which is characteristic of the quantized relativistic particle.
More precisely, we have the uncertainty relation:
$$
(\delta X_\mu)(\delta X_\nu)\,\geq\, \f{l_C^2}{2}\,|\la
\what{\jmath}_\mn\ra|.
$$
Let us look at the space sector and consider the uncertainty in
the spatial position, $(\delta l)^2\,\equiv\,(\delta
X_1)^2+(\delta X_2)^2+(\delta X_3)^2$. Following arguments from
\cite{dan2}, one shows that $\delta l$ is always larger than the
Compton length $l_C$ as long as the state is not invariant under
$\SO(4)$ i.e under the $J_{ij}$'s and the $X_i$'s. However, if the
state is invariant under $\SO(4)$, it can not be invariant under
$X_0$ and the uncertainty $(\delta X_0)$ will be larger than
$l_C$. A more explicit analysis would require more details on the
action of the $\so(4,1)$ generators. However, from the study of
$\so(3,1)$ in \cite{dan2}, we expect that this would naturally
lead to a position uncertainty always larger than $l_C$.

\section{Quantization on the 5d light-cone}\label{5dquant}
We now proceed to the quantization of the algebra of observables for a fixed mass $m>0$. We will
see that the observables $j,X(T),p$ can naturally be represented on the five-dimensional light
cone. To this purpose, let us introduce the 5d coordinates $y_A$ and their conjugate momentum
variables $\pi_A$, with the symplectic structure $\{y_A,\pi_B\}=\eta_{AB}$ and the metric
$\eta_{AB}=(+----)$. We define the 5d light cone $\cc_0$ in momentum space as the algebraic
manifold:
\be
\cc_0 \equiv \{(\pi_A) \,|\, \pi_0^2-\pi_i\pi_i-\pi_4^2=0\}.
\ee
Choosing a particular value of the time $T_0$, we identify $j$ and $X(T_0)$ to the 5d Lorentz generators,
$$
J_{AB}=y_A\pi_B-y_B\pi_A,
$$
and we will define the 4-momentum $p_\mu$ as a simple function of the 5d momenta $\pi_A$,
\be
p_\mu\equiv m\f{\pi_\mu}{\pi_4},
\ee
so that the mass-shell condition $p^2=m^2$ becomes the {\it light-cone condition},
\be
p_\mu p^\mu-m^2\,=\,\f{m^2}{\pi_4^2} \pi_A\pi^A.
\ee
It is straightforward to check that this choice has the right Poisson bracket with $X_\mu(T_0)$
and $j_\mn$. This maps 4d massive relativistic particles to 5d massless particles (at the level of
Dirac observables). This provides us with a natural 5-dimensional action principle for the 4d
relativistic particle. We will analyze this in details in the next section.

At the quantum level, we will work in the $y$-polarisation. We represent the 5-momentum $\pi$ as
derivation operators,
$$
\what \pi_A\,=\,
-i\eta_{AB}\f{\pp}{\pp y_B},
$$
and the observables $X_\mu(T_0)$ becomes the differential operators $\what J_{AB}/m$.

Next we would like to identify the reference time $T_0$ to the quadratic Casimir $J_{AB}J^{AB}$. At
the classical level, we have\footnotemark:
\be
\f{1}{2}J_{AB}J^{AB}=(y_Ay^A)(\pi_B\pi^B)-(y_A\pi^A)^2,
\ee
where we introduce the 5d dilatation $\dd=y_A\pi^A$.
\footnotetext{
$(y_Ay^A)(\pi_B\pi^B)-(y_A\pi^A)^2$ is actually the quadratic Casimir of the $\sl(2,\R)$ Lie
algebra generated by the operator $\dd=y_A\pi^A$, $(y_Ay^A)$ and $(\pi_A\pi^A)$.} Since $\{y_Ay^A,
\pi_B\pi^B\}=4\dd$, we only expect linear terms in $\dd$ due to
ordering ambiguities at the quantum level. Introducing the dilatation and laplacian operators,
$$
\what{\dd}=y_A\pp^A,\quad
\Delta=\pp_A\pp^A,
$$
we can compute the Casimir operator $\what J^2$ for the algebra $\so(4,1)$ (we are working at $d=4$): \bes -\f{1}{2}\what
J_{AB}\what J^{AB}&=&
y^2\Delta- \what{\dd}(\what{\dd}+d-1) \\
&=&\Delta y^2-\what{\dd}(\what{\dd}+d+3)-2(d+1)\nn \\
&=&\f12(y^2\Delta+\Delta y^2)-\what{\dd}(\what{\dd}+d+1)-(d+1).\nn
\ees
The extra term arises from the ordering ambiguity since $\Delta$ and $y^2$ do not commute. This is
analogous to the extra factor met for example in the vacuum energy of the harmonic oscillator. Note
that $\what{\dd}$ (and its square) is not (anti-)Hermitian, and we actually have:
$$
\what{\dd}\dag\,=\,-\what{\dd}-(d+1).
$$
The shifts in the operator $\what J_{AB}\what J^{AB}$  ensure that the Casimir operator remains
self-adjoint.

We would like to compute the eigenvalues of ${\what J}^2$ on the light cone $\cc_0$ i.e on the states
$\vphi$ satisfying $\Delta \vphi=0$. We introduce the states $\vphi_{P,\lambda}(y)=(y_A
P^A)^\lambda$. They satisfy $\Delta\vphi_{P,\lambda}=0$ as soon as $P$ is light-like, $P^AP_A=0$.
Moreover they diagonalise the dilatation operator,
$\what{\dd}\vphi_{P,\lambda}=\lambda\vphi_{P,\lambda}$. Hence, we have:
\be
-\f{1}{2}\what J_{AB}\what J^{AB} \,\vphi_{P,\lambda}\,=\,
-\lambda(\lambda+d-1)\,\vphi_{P,\lambda}.
\ee
We therefore identify the time to the eigenvalue, $T_0^2=-\lambda(\lambda+d-1)$. This is possible
if and only if $\lambda$ has a fixed real part:
\be
\lambda=-\f{(d-1)}{2}+i\beta,\qquad
T_0^2=\beta^2+\left(\f{d-1}{2}\right)^2.
\ee
The time remains continuous at the quantum level, even though we have a minimal time unit
$T_{min}=3/2$ (in Compton unit $\hbar/mc^2$).

To summarize, the space of harmonic functions (having a vanishing laplacian) form a reducible representation of $\so(4,1)$ which
decomposes into irreducible representations labeled by the parameter $\lambda$. These irreducible components are formed by
homogeneous functions and $\lambda$ is the corresponding eigenvalue of the dilatation operator. Then the time $T_0$ actually
fixes which representation we use.

Finally, we can check that since $p_\mu, j_\mn, X_\mu$ all commute with the 5d dilatation
$\what{\dd}$, it is natural that we can represent them in an irreducible representation
corresponding to a single eigenvalue of $\dd$.

\medskip

Now, we are interested in the precise mapping between the usual 4d wave functions and the states of
our 5d quantization. For this purpose, it is easier to use the momentum polarisation and work with
the Fourier transforms. Let us introduce the following family of morphisms between the usual space
of wave functions $\vphi(p_\mu)$ and the space of functions on the light cone:
\be
\Theta_\alpha:\Phi(p_\mu)\,\arr\,\phi(\pi_A)=(\pi_4)^{\alpha}\,\Phi\left(m\f{\pi_\mu}{\pi_4}\right).
\ee
We use the Fourier transform on $\cc_0$:
$$
\vphi(y)=\int d^5\pi\,
\delta(\pi_A\pi^A) e^{iy_A\pi^A}\phi(\pi).
$$
Using this Fourier transform, it is straightforward to check that $\Theta_\alpha$ gives functions
which are eigenvectors of $\hat{\dd}$ with eigenvalue\footnotemark:
$$
\lambda=-(d-1)-\alpha.
$$
\footnotetext{
If we had not include the $\delta(\pi^2)$ in the measure of the Fourier transform, we would have
found $\lambda=-(d+1)-\alpha$. This shift $(d+1)\arr(d-1)$ is exactly the same as above in the
analysis of the Hermicity of $X_\mu$ with respect to the kinematical and physical inner product.}

Using the isomorphism $\Theta_\alpha$, we can compute the action of the operator $\what X_\mu=
J_{\mu 4}/m$ on the standard wave functions $\Phi(p)$. Working in the $\pi$-polarisation, $\what{y}$
acts as $+i\pp/\pp\pi$ and $X_\mu$ becomes:
$$
\what X_\mu=+\f{i}{m}\left( \pi_\mu\f{\pp}{\pp \pi_4}+\eta_{\mu\nu}\pi_4\f{\pp}{\pp \pi_\nu} \right).
$$
Then defining $\what X_\mu^{(\alpha)}\equiv \Theta_\alpha^{-1}\what X_\mu \Theta_\alpha$, we obtain: \be \what
X_\mu^{(\alpha)}\phi(p)=+i\eta_{\mu\nu}\f{\pp\phi}{\pp p_\nu} -i\f{p_\mu p_\nu}{m^2}\f{\pp\phi}{\pp p_\nu}
+i\alpha\f{p_\mu}{m^2}\phi. \label{Xwhat} \ee Computing the action of this operator on standard plane waves
$\vphi_x(p)=e^{-ip.x}$, we get: \be \what X_\mu^{(\alpha)}\vphi_x =\left(x_\mu+\f{p_\mu}{m^2}(+i\alpha-p_\nu
x^\nu)\right)\vphi_x. \ee Inserting $+i\alpha=-i(d-1)-i\lambda=+\beta-i(d-1)/2$ in the previous formulae, we recognize the exact
same equation as in \Ref{timeshift} with the same imaginary shift in time $-i(d-1)/2$. This imaginary shift is purely a quantum
effect.

At the end of the day, the shift due to the ordering ambiguities in the 5d quantization fits
exactly the shift required for the hermiticity of the observables $X_\mu(T)$ with respect to the
physical inner product: the 5d quantization on the light cone is equivalent to the standard
quantization of the 4d relativistic particle.

Finally, we point out that when $\lambda$ is set to zero, we recover the representation of Snyder's non-commutative coordinates
as differential operators in the momentum $p$ \cite{snyder}. Notice nevertheless that  $\lambda=0$ is actually excluded in our
analysis due to the quantum shift in $-i(d-1)/2$.

\section{A 5d action principle and DSR}

\subs{From 5d to 4d}

Since we represent the algebra of observables of the relativistic particle on the 5d
light cone, it seems natural to propose the following 5d action principle for the massive
4d particle as a massless 5d particle:
\be
S_{5d}=\int \, \pi^A {\rm d}y_A -\lambda\,\pi_A\pi^A.
\ee
We have the 5d mass-shell condition $\hh_{5d}=\pi_A\pi^A$ and the mass $m$ of particle
does not appear in this 5d action. The natural issue is how to recover the standard
relativistic particle described in term of $(x_\mu,p_\mu)$.

Following the previous section, we introduce the variables:
\be
p_\mu=m\f{\pi_\mu}{\pi_4},\quad
X_\mu=\f{1}{m}\left(y_\mu\pi_4-y_4\pi_\mu\right),
\ee
where $m\in\R_+^*$ is an arbitrary fixed parameter. They are both
Dirac observables, $\{\hh_{5d},X\}=\{\hh_{5d},p\}=0$ and we check
their Poisson brackets:
$$
\{X_\mu,p_\nu\}=\eta_\mn-\f{p_\mu p_\nu}{m^2},\quad \{X_\mu,X_\nu\}=-\f{j_\mn}{m^2},
$$
with the Lorentz generators $j_\mn=y_{[\mu}\pi_{\nu]}=X_{[\mu}p_{\nu]}$.

Expressing $y_\mu$ in terms of the new variables, the 5d kinetic term reads as, up to a total derivative: \be \pi^A{\rm d}y_A=
p^\mu{\rm d}X_\mu+\ln\pi_4{\rm d}\left(p^\mu X_\mu\right) -\f{1}{2}\f{y_4}{\pi_4}{\rm d}\left(\pi^A\pi_A\right). \ee On-shell,
$\pi^A\pi_A$ is fixed (to zero) and the last term vanishes. The first term is the usual 4d kinetic term stating that $p_\mu$ and
$X_\mu$ are conjugate momenta/positions. The second term involves $p^\mu X_\mu$, which is our (Lorentz-invariant) clock time $T$
measuring the proper time along the particle's trajectory in the usual 4d space-time. This identifies $\ln\pi_4$ as the conjugate
momentum to the proper time: it generates the Hamiltonian evolution. This should be compared to the (effective) Hamiltonian $\ln
H$ describing the evolution the (weak) observables $X$ as written in equation \Ref{Heff}. Finally, we can interpret $y_4/\pi_4$
as the conjugate coordinate to the 5d mass $\pi^A\pi_A$.

Up to now, we have dealt with the position observables $X_\mu$. It would be interesting
to recover the standard commutative 4d space-time coordinates. We first notice that:
$$
p^\mu X_\mu= \pi^A y_A -\f{y_4}{\pi_4}\pi^A\pi_A.
$$
On the 5d mass-shell, $\pi^A\pi_A=0$, fixing $p^\mu X_\mu=T$ is thus equivalent to fixing
$\dd=\pi^A y_A=T$. Assuming this extra condition, $\dd=T$, the coordinate $X_\mu$ is
actually exactly the Dirac observables for the relativistic particle that we introduced
earlier \Ref{Xobs}. More precisely, we introduce 4d coordinates as:
\be
x_\mu=y_\mu \f{\pi_4}{m}.
\ee
It is straightforward to check that they satisfy the standard canonical Poisson brackets:
$$
\{x_\mu,x_\nu\}=0, \quad
\{x_\mu,p_\nu\}=\delta_\mn.
$$
Moreover we now have:
$$
\left|
\begin{array}{c}
\hh_{5d}=0\\
\dd=T
\end{array}
\right. \,\Rightarrow X_\mu=x_\mu+\f{p_\mu}{m^2}(T-x.p)=X_\mu(T).
$$
Finally, writing the 5d kinetics in terms of $x_\mu$, we get (up to a total derivative): \be \pi^A {\rm d}y_A=p^\mu {\rm d}x_\mu
+ \ln\pi_4\,{\rm d} \left(\pi^A y_A\right). \ee In light of these remarks, we propose to reduce the 5d system to a 4d one through
a gauge fixing of the 5d mass-shell condition $\hh_{5d}$. We choose as gauge fixing condition $\dd=T$. It is straightforward to
compute the corresponding Dirac bracket: \bes
\{y_A,\pi_B\}_D &=& \eta_{AB}-\f{\pi_A\pi_B}{\pi^C\pi_C},\nn\\
\{y_A,y_B\}_D &=& -\f{J_{AB}}{\pi^C\pi_C}.\label{Dirac5d}
\ees
Since $p_\mu$ and $X_\mu$ commute with both $\hh_{5d}$ and the 5d
dilatation $\dd$, their Dirac bracket with any phase space
function is equal to the Poisson bracket. Then we check that
$\ln\pi_4$ generates the Hamiltonian flow on the 4d variables:
\bes
&&\{\ln\pi_4, X_\mu\}_D=\{\ln\pi_4, p_\mu\}_D=0,\nn\\
&&\{\ln\pi_4, x_\mu\}_D=\,p_\mu\,\f{\pi_4^2}{m^2\pi^A\pi_A}.
\label{pi4flow}
\ees
The usual 4d Hamiltonian constraint $\hh_{4d}=p^2-m^2$ is easily expressed in terms of the 5d variables:
$$
\hh_{4d}=m^2\f{\pi^A\pi_A}{\pi_4^2}.
$$
Its flow is thus the same as the one generated by $\pi_4$ up to a factor $\pi^A\pi_A$. We check that it of course removes all the
$\pi_4$ and $\pi^A\pi_A$ factors and simply we recover the usual relation $\{\hh_{4d},x_\mu\}_D=-2p_\mu$.

Therefore, if we want to recover the standard relativistic
dynamics after gauge fixing, we have to add an extra-constraint to
the 5d action and we write:
\be
S_{5d}=\int \, \pi^A {\rm d}y_A-\lambda\,\pi^A\pi_A -\mu(\pi_4-M),
\ee
where $M$ is an arbitrary parameter. The path integral for this
action is obviously equivalent to a relativistic particle with
action $\int \pi^\mu {\rm d}y_\mu-\lambda(\pi^\mu\pi_\mu-M^2)$.
This shows that it is the fifth moment $\pi_4$ which generates the
(rest) mass of the 4d particle.

\subs{DSR as a regularization}

The main problem with the gauge fixing procedure is the singularity of the Dirac bracket at $\pi^A\pi_A=0$ on the 5d mass-shell.
A natural regularization is to allow a (small) deviation from 0 and modify the 5d constraint to: \be
\hh_{5d}=\pi^A\pi_A+\epsilon\kappa^2=0, \ee where $\kappa\in\R_+^*$ is a mass scale and $\epsilon=\pm$ the sign of the deviation.
The full 5d action now reads: \be S_{5d}=\int \, \pi^A {\rm d}y_A-\lambda\,(\pi^A\pi_A+\eps\ka^2)-\mu(\pi_4-M), \ee This is
actually the 5d action which generates DSR. Indeed, it has been shown in \cite{GKKL} that all the various bases of Deformed
Special Relativity can be derived as different gauge fixing of the 5d constraint $\hh_{5d}$. In that framework, $\eps$ is
required to be positive and the momentum space $\pi^A\pi_A=-\ka^2$ is the de Sitter space. $\ka$ is usually set to the Planck
mass and induces a discrete spectrum for certain distance operators \cite{dan1}. The rest mass of the 4d particle is obtained
from the $\pi_4$ constraint and is a function of $M$ (and $\ka$), the exact function depending of the details of the gauge
fixing.

Having $\ka\ne 0$ regulates all the previous expressions. Moreover
it allows to explicitly invert the definition of the 4-momenta
$p_\mu=m\pi_\mu/\pi_4$ and express the fifth moment $\pi_4$ in
term of $p^2$:
\be
\pi_4^2=\f{\eps\ka^2}{1-\f{p^2}{m^2}}.
\ee
For $\eps=+$, we are constrained to work with a bounded momentum
$p^2\le m^2$. The other choice $\eps=-$ leads to $p^2\ge m^2$ and
we discard it as unphysical. Since $\pi_4$ is a simple function of
$p^2$, it is now obvious that it generates the 4d Hamiltonian flow
for the relativistic particle. Moreover, the parameter $m$ loses
its straightforward interpretation as the rest mass of the
particle. More precisely, imposing the constraint $\pi_4=M$, we
obtain:
$$
p^2=m^2\left(1-\f{\ka^2}{M^2}\right).
$$
We recover the standard dispersion relation $p^2=m^2$ in the limit
$\ka\ll M$ when we remove the regulator $\ka\arr0$.

This shows how classical mechanics in DSR can be considered as a regularization of standard Special Relativity from the
five-dimensional point of view when quantizing the algebra of Dirac observables. This is consistent with the hope that DSR
quantum field theory regularizes Feynman diagrams of standard QFT. This offers a shift of perspective on the interpretation of
the non-commutative space-time coordinates of DSR. Indeed the non-commutativity of Lorentz-covariant position observables is
already present in Special Relativity (consider the $X_\mu(T)$ coordinates). These coordinates are not the true space-time
coordinates, $x_\mu$, but Dirac observables which are constants of motion. This is supported by the fact that the relativistic
Dirac observables $X_\mu,p_\mu$ have the same Poisson brackets as the DSR phase space coordinates in the Snyder basis
\cite{snyder}. The only difference is that we represent the Poisson algebra of the 5d light cone while the DSR coordinates are
realized as operators on the de Sitter space \cite{snyder,DSR}.

\subsection{A 5d representation of the Feynman propagator}

Since we have provided a 5d representation of the relativistic particle at both the level
of the action and of the Dirac observables, it would be interesting to investigate
whether this picture can be extended to quantum field theory. More precisely, we focus on
the Feynman propagator, from which one can then build the whole perturbative expansion of
the scattering amplitudes.

In the proper time representation, the Feynman propagator reads:
\be
K_m(x_\mu)\,\equiv\,
\int_{R_+} dT \int d^4p_\mu\, e^{ip^\mu x_\mu}
e^{iT(p^2-m^2+i\eps)},
\ee
where $\eps>0$ is a regulator. We would like to express this in terms of the 5d variables
$(y_A, \pi_A)$. First, we write the mass-shell constraint in terms of the $\pi$'s:
$$
p^2-m^2\,=\, m^2\f{\pi_A\pi^A}{\pi_4^2}.
$$
We perform a first change of variable $p_\mu=m\pi_\mu/\pi_4$. It is then natural to
introduce the rescaled coordinates $y_\mu=m x_\mu/\pi_4$ to preserve the symplectic form. Finally, we do a change of
variables from the proper time $T$ to the fifth coordinate $y_4=T\,m^2\pi_A\pi^A/\pi_4^3$.
We obtain in the end:
\be
K_m(x)\,=\, m^2\,\int_{\R_+} dy_4 \int \f{d^4\pi_\mu}{\pi_4}\,
\f{e^{i\pi^A y_A}}{\pi_A\pi^A}.
\ee
A first remark is that $\pi_4$ is still unspecified. It could possibly play the role of a
renormalisation scale or some energy cut-off. Nevertheless, the measure
$d^4\pi_\mu\,/\pi_4$ suggests a lift to a 5d integral such as
$d^5\pi_A\,\delta(\pi_A\pi^A)$. However, this would conflict with the $1/\pi_A\pi^A$ term
in the integral. This is normal since imposing $\pi_A\pi^A=0$ amounts to enforcing the
mass-shell constraint, but the Feynman propagator is an off-shell object.

\medskip

We propose to resolve this issue by the same DSR regularization as used earlier. We
introduce the constraint $\delta(\pi_A\pi^A+\ka^2)$. Then we obtain the following 5d
representation of the Feynman propagator:
\be
K_m(x)
\,=\,
\f{m^2}{\ka^2}\,\int_{\R_+} dy_4 \int d^5\pi_A\,\delta(\pi_A\pi^A+\ka^2)\,
e^{i\pi^A y_A}.
\ee
We have thus written the Feynman propagator as an on-shell object from the 5d point of
view.

The only subtle point is that imposing $\pi_A\pi^A+\ka^2=0$ truncates the momentum space
to the $p^2<m^2$ sector. To recover the other half of the momentum space, we should switch
the sign of $\ka^2$ and impose $\pi_A\pi^A-\ka^2=0$. Therefore the full Feynman
propagator, with an integration over the whole $p$ space, is the difference of the two 5d
integrals with $\ka$-shell condition respectively $\delta(\pi_A\pi^A+\ka^2)$ and
$\delta(\pi_A\pi^A-\ka^2)$.

This 5d representation of the Feynman propagator allows a clear interpretation of the
fifth dimension: $\pi_4$ represents the mass of the particle (or more precisely the
possibly off-shell $p^2$) while $y_4$ is the proper time (rescaled by some $m/\pi_4$
factor).

\medskip

This 5d reformulation of the Feynman propagator should allow a 5d DSR-like representation
of all scattering amplitudes of quantum field theory. From the reverse point of view, it
shows that amplitudes computed in QFT based on DSR could simply be equivalent to standard
QFT. To evade such a no-go theorem about DSR, we see two alternatives:
\begin{itemize}
\item DSR relaxes the $\ka$-shell condition, either by allowing
$\ka$ to vary or by allowing $\pi_A \pi^A$ not to be fixed at $\pm
\ka^2$. In this case, DSR will truly be a 5d theory based on a physical 5d
momentum $\pi_\mu$.
\item The deformation of the scattering in DSR is not strictly
contained in the Feynman propagator but is due to a modification of the interaction
vertices. In algebraic terms, we deform the co-product dictating the law of addition of
the moments \cite{FKN}.
\end{itemize}
These two viewpoints do not exclude each other.

\section{Ordered observables and Feynman diagrams}

Other useful relational observables are the ones recording whether
the particle went through a given fixed space-time point $z_\mu$
along its trajectory. Let us introduce the phase space
distribution:
\be
\oo_z\equiv \int_\R d\tau\, \delta^{(4)}\left(x_\mu\left(\f\tau m\right)-z_\mu\right),
\ee
where $x_\mu(\tau)=\exp(-\tau\{H,\cdot\}/2)\,x_\mu=x_\mu+\tau p_\mu$. The factor $m$ is here for dimensional purposes.
It is straightforward to check that this is a Dirac observable.
Carrying out the integration in $x_0$, this observable reads as:
\be
\oo_z=\f{m^2}{p_0^2}\delta^{(3)}(K_i(z)),
\ee
where we have defined the boost vector $K_i(z)\equiv
p_0(x_i-z_i)-p_i(x_0-z_0)$. Note that
$K_i(z)=j_{0i}-(p_0z_i-p_iz_0)$. We can easily compute the Fourier
transformed observable:
\be
\oo_q\equiv\int d^4z e^{iq.z}\oo_z = e^{iq.x}\delta\left(\f{q.p}{m^2}\right).
\ee
We quantize this operator by splitting the $\exp(iq.x)$ into two and we define:
\be
\what{\oo}_q\,\ket{p}=\delta\left(\f{\left(p+\f q2\right).q}{m^2}\right)\,\ket{p+q}.
\ee
Due to the chosen ordering, this operator is still a Dirac
observable at the quantum level, i.e it commutes with the quantum
operator $p^2$ and leaves invariant the Hilbert of physical state
(annihilated by $\what{p}^2-m^2$). Indeed, the $\delta$-function,
$$
m^2\delta\left(\left(p+\f q2\right).q\right)=
2m^2\delta\left((p+q)^2-p^2\right),
$$
imposes that $(p+q)$ is on the same mass-shell than $p$.
Finally, reversing the Fourier transform, we define the quantum operator:
\be
\what{\oo}_z\,\ket{p}=\int_\R d\tau\int dq\, e^{-iq.\left(z-\tau\f{p+\f q2}{m^2}\right)}\,\ket{p+q}.
\ee
The $\f q 2$ shift in the exponential is due to the quantum  ordering.

An interesting related phase space function is defined by restricting the range of $\tau$-integration to $\R_+$:
\be
\ff_z=\int_{\R_+}d\tau\, \delta^{(4)}\left(x_\mu\left(\f\tau m\right)-z_\mu\right).
\ee
Taking the Fourier transform, we define:
\be
\ff_q=\int d^4z\,e^{iq.z}\ff_z=e^{iq.x}\f{im}{q.p+i\eps},
\ee
where $\eps>0$ regularizes the $\tau$-integration. The $\ff$'s are
not Dirac observables but their Poisson bracket with the
Hamiltonian constraint generates the plane waves\footnotemark:
\be
\{H,\ff_q\}=2me^{iq.x}.
\ee
\footnotetext{We also compute the bracket of the plane waves with $\ff$:
$$
\{e^{ir.x},\ff_q\}=\f{m(r.q)e^{ix.(q+r)}}{(q.p+i\eps)^2}
=\f{r.q}{m}e^{-ix.q}\ff_q^2.
$$}
We quantize $\ff_q$ using the same ordering as for $\oo_q$ splitting the translation $e^{iqx}$ into two halves:
\bes
\what{\ff_q}\,\ket{p}&=&\f{im}{q.\left(p+\f q 2\right)+i\eps}\,\ket{p+q},\\
&=&\f{2im}{(p+q)^2-p^2+i\eps}\,\ket{p+q}.
\ees
Its commutator with the Hamiltonian generates simple translations in the momentum space:
\be
\left[\what{H},\what{\ff_q}\right]\,\ket{p}
\,=\, 2im\,\ket{p+q}.
\ee

It is straightforward to generalize to observables recording
whether the particle went through a certain number of space-time
points, $z^{i}$, ordered in time:
$$
\int \prod_{i=1}^n{\rm d\tau_i}\,
\delta^{(4)}(x_\mu(\frac{\tau_i}{m})-z^{i}_\mu).
$$
We define $\oo^{(n)}_{z^i}$ for a range
$-\infty<\tau_1<..<\tau_n<+\infty$ and $\ff^{(n)}_{z^i}$ for the
restricted range $0<\tau_1<..<\tau_n<+\infty$.
We compute their Fourier transform, now depending on $n$ momenta $q^i$:
$$
\oo^{(n)}_{q^i}=\f{m^n\delta(p.Q_1)e^{ix.Q_1}}{\prod_{j=2}^n (p.Q_j+i\eps)},
\quad
\ff^{(n)}_{q^i}=\f{(im)^ne^{ix.Q_1}}{\prod_{j=1}^n (p.Q_j+i\eps)},
$$
$$
\oo^{(n)}_{q^i}=\f{m}{i^{n-1}}\delta(p.Q_1)e^{ix.q^1}\ff^{(n-1)}_{q^2,..,q^{n}},
$$
where we have defined the momenta $Q_j\equiv\sum_{i=j}^n q^i$.
$\oo_{q^i}$ is obviously once again a Dirac observable. As for the
$\ff$'s, it is straightforward to compute:
\be
\{H,\ff^{(n)}_{q^1,..,q^n}\}
\,=\,
2m e^{ix.q^1}\,\ff^{(n-1)}_{q^2,..,q^n}.
\ee
At the quantum level,we similarly define the operators:
\be
\what{\oo}^{(n)}_{q^i}\,\ket{p}
=\f{2m^n\,\delta((p+Q_1)^2-p^2)}{\prod_{j=2}^n \left(Q_j.\left(p+\f{Q_j}{2}\right)+i\eps\right)}\,\ket{p+Q_1},
\ee
\be
\what{\ff}^{(n)}_{q^i}\,\ket{p}
=\f{(im)^n}{\prod_{j=1}^n \left(Q_j.\left(p+\f{Q_j}{2}\right)+i\eps\right)}\,\ket{p+Q_1}.
\ee
This leads to the same tower of operators with the following commutators:
$$
[\what{H},\what{\ff}^{(n)}]\,=\,
2m\,T_{q^1}\,\what{\ff}^{(n-1)},
$$
where $T_q$ is the momentum translation by $q$.

\medskip

The one-loop Feynman diagram with two external legs can be extracted from:
\be
\int d^4s\, \la p\,|\what{\oo}^{(4)}_{q,s,r,-s}|p\ra,
\ee
where $p$ is a reference vector on the mass-shell. Indeed up to the $\eps$
regulator, this is equal to:
$$
(2m)^4\delta^{(4)}(q+r)
\f{\delta((p+q+r)^2-p^2)}{(p+r)^2-p^2}\,\ii_2(r),
$$
$$
\ii_2(r)=\int d^4s\,\f{1}{((r+s)^2-p^2)(s^2-p^2)}.
$$
Setting $p^2=m^2$, we obtain the one-loop Feynman amplitude with two legs for a
massive scalar field, up to a normalization factor $1/((p+r)^2-p^2)$. We can
further get rid of the $\delta((p+q+r)^2-p^2)$ which leads to a singular result
by equivalently considering the operator $\exp(ix.q)\ff^{(3)}_{s,r,-s}$:
\be
\int d^4s\, \la p\,|\what{e^{ix.q}}\what{\ff}^{(3)}_{s,r,-s}|p\ra
=
\delta^{(4)}(q+r)
\f{(2im)^3}{(p+r)^2-p^2}\,\ii_2.
\ee
Note that despite the fact that we are using $\ff$, that operator still defines a Dirac
observable since $\oo^{(n)}\sim \exp(ix.q)\ff^{(n-1)}$.

The physical interpretation of $\oo^{(4)}_{q,s,r,-s}$ is as
follows. Since we are dealing with $\oo^{(4)}$, we are
constraining the particle to go through four fixed space-time
points {\it ordered in time along the particle trajectory}.
However, due to the identification of the moments
$s\leftrightarrow -s$, the two last points are actually identified
to the two first points, so that the particle does a time-like
loop, going through a point $z_1$ then through a point $z_2$ then
point $z_1$ again and point $z_2$ again. This graph exactly draws
the one-loop Feynman diagram with two external legs.

This can be generalized to all Feynman diagrams. This shows that the quantum field theory
scattering amplitudes can be expressed as expectation values of the $\oo^{(n)}$ Dirac observables
(for a single relativistic particle).

Let us conclude this section by the following remark. If we were to define the
$\oo$ observables using the Dirac observables $X_\mu(\tau)$ instead of the
coordinates $x_\mu(\tau)$,
$$
\tl{\oo}_z\equiv \int_\R d\tau\, \delta^{(4)}\left(X_\mu\left(\f\tau m\right)-z_\mu\right),
$$
its Fourier transform $\tl{\oo}_q$ would not change at all, \be \tl{\oo}_q\equiv\int d^4z e^{iq.z}\tl{\oo}_z =
e^{iq.x}\delta\left(\f{q.p}{m^2}\right)=\oo_q, \ee and the whole following construction would be identical. It is thus possible
to construct the Feynman diagram evaluations from expectation values of the $X_\mu(T)$ Dirac observables. We believe this can be
applied in order to define the Feynman amplitudes for DSR. Following \cite{eteralaurent}, the main ingredient would be a modified
Fourier transform reflecting that the momentum space is not the flat Minkowski space but is not curved.

\section{The Particle with Spin}

\subs{Dirac observables}

In this final section, we investigate if including spin will
change the analysis of the algebra of Dirac observables of the
relativistic particle. A massive spinning particle is defined by
the phase space $(x_\mu,p_\mu,s_\mn)$. The Poisson bracket still
defines $x$ and $p$ as canonical variable. The $s_\mn$ commute
with $x$ and $p$ and form a Lorentz algebra, i.e. have similar
brackets as $j_\mn$. The constraints on the phase space are now:
\bes
H&=&p^2-m^2 \\
N&=&\f{1}{2}s_\mn s^\mn +\lambda^2\\
O^\nu&=& p_\mu s^\mn,
\ees
where $\lambda$ is the norm of the spin. This formalism can be entirely derived from a Lagrangian\footnotemark.
\footnotetext{
We consider a matrix $\Lambda\in \SO_o^{\uparrow}(1,3))$ and define the momentum and the spin as:
$$
p_\mu=m\Lambda_{\mu 0}, \qquad
\f{1}{2}s_\mn J^\mn=-is =\lambda\Lambda^{-1}J_{12}\Lambda,
$$
where $J_\mn$ are the Lorentz generators in the fundamental representation in term of
$4\times 4$ matrices. It is easy to check that $s$ explicitly reads as:
$$
s_\mn=\lambda(\Lambda_{\mu 1}\Lambda_{\nu 2}-\Lambda_{\mu 2}\Lambda_{\nu 1}).
$$
Then we introduce the action:
$$
{\cal S}=\int d\tau\, p^\mu\dot{x}_{\mu}
-i\f{\lambda}{2}\textrm{Tr}(J_{12}\Lambda^{-1}\dot{\Lambda}).
$$
It is straightforward to check that the spin term in the
Lagrangian is $\f{\lambda}{2}(\Lambda_{\mu 2}\dot{\Lambda}_{\mu
1}-\Lambda_{\mu 1}\dot{\Lambda}_{\mu 2})$, so that the resulting
symplectic structure is simply $\{\Lambda_{\mu 1},\Lambda_{\mu
2}\}=1/\lambda$. From there, it is obvious that the $s$'s form a
Lorentz algebra under the Poisson bracket. }

Dirac observables now need to commute with all 6 constraints. The algebra of observables is now  generated by the momentum
$p_\mu$ and the Lorentz generators $L_\mn=j_\mn+s_\mn$. We actually recover the Poincar\'e algebra. Let us notice that the spin
$s_\mn$ is not an observable although it is a constant of the motion (since it commutes with the Hamiltonian constraint $H$
generating the trajectories and doesn't with $O_\mu$).

It is still straightforward to construct the coordinate Dirac
observables. And we define the position observables in term of a
time $x\cdot v$:
\be
X_\mu^{(v)}=\f{L_{\mu\nu}v^\nu+Tp_\mu}{p\cdot v}
=x_\mu+\f{p_\mu}{p\cdot v}\left(T-x\cdot v\right)+\f{s_\mn v^\nu}{p\cdot v},
\ee
and in term of the time $D$:
\be
X_\mu=\f{L_{\mu\nu}p^\nu+Tp_\mu}{p^2}
=x_\mu+\f{p_\mu}{p^2}(T-D)+\f{O_\mu}{p^2},
\ee
which is weakly equal to the spinless case. Similarly to  the spinless case, we obtain:
$$
\{X_\mu,X_\nu\}=-\f{L_\mn}{p^2}=-\f{j_\mn}{p^2}-\f{s_\mn}{p^2}.
$$
The spin $s$ creates an additional non-commutativity.

Another useful Dirac observable is the Pauli vector:
$$
W_\mu=\epsilon_{\mu\nu\lambda\rho}p^\nu L^{\lambda\rho}
=\epsilon_{\mu\nu\lambda\rho}p^\nu s^{\lambda\rho}.
$$
$W_\mu$  behaves as a vector under Lorentz transformations. Its
norm $W_\mu W^\mu$ is the second Casimir of the Poincar\'e
algebra: $W_\mu W^\mu=m^2\lambda^2$. We have the following Poisson
brackets:
$$
\{W,p\}=0,\qquad
\{W_\mu,X_\nu\}=\f{p_\mu W_\nu}{p^2}.
$$
Let us also point out the identity:
$$
O_\mu W^\mu= \f{1}{4}p^2 \epsilon^{\alpha\beta\gamma\delta}s_{\alpha\beta}s_{\gamma\delta}.
$$

Next, we can introduce Dirac observables for the spin. As $s_\mn$
commutes with $H$ and $N$, we only need to take care of the
constraint $O_\mu$ and gauge fix it. Thus we introduce the
following Dirac observables labelled by an arbitrary fixed {\it
vector} $a_\mu$:
\bes
S_\mn^{(a)}&=&L_\mn-(a_\mu p_\nu-a_\nu p_\mu)\nn\\
&=&s_\mn + (x_\mu-a_\mu) p_\nu - (x_\nu-a_\nu) p_\mu,
\ees
which gives $s_\mn$ when $x_\mu=a_\mu$. These observables
$S_\mn^{(a)}$ satisfy the same brackets as $s_\mn$ and also form a
Lorentz algebra. Moreover $S_\mn^{(a)}$ act as the Lorentz
generators on the momentum vector $p_\mu$.

We can remove the dependence on an arbitrary fixed vector $a_\mu$ and render the expression covariant by using the
Dirac observable $X_\mu$. Thus we introduce our spin Dirac observable:
\bes
S_\mn&=&L_\mn-(X_\mu p_\nu-X_\nu p_\mu)\nn\\
&=& s_\mn + (x_\mu-X_\mu) p_\nu - (x_\nu-X_\nu) p_\mu \nn\\
&=& s_\mn +  \f{1}{p^2}\left(O_\mu p_\nu -  O_\nu p_\mu\right),
\ees
which is actually weakly equal to the spin $s_\mn$ itself. Moreover $\{S,p\}=0$ and the Poisson brackets of the $S$'s
form the Lorentz algebra on-shell (off-shell, we get a few $O\wedge p$ terms).

\subs{Spin-induced non-commutativity}

One can compute the algebra of constraints of the relativistic spinning particle and we found:
$$
\{H,O_\mu\}=\{N,O_\mu\}=\{H,N\}=0,
$$
\be
\{O_\mu,O_\nu\}=p^2 s_\mn +(O_\mu p_\nu - O_\nu p_\mu)=p^2 S_\mn.
\ee
So $H$ and $N$ are first class constraints, while the $O_\mu$'s are second class
constraints. One can thus introduce the Dirac bracket taking the constraints $O_\mu=0$
into account. Noting $\Delta_\mn\equiv \{O_\mu,O_\nu\}$ the Dirac matrix, we can compute
its inverse:
\be
(\Delta^{-1})^\mn=\f{-\eps^{\mn\alpha\beta}\Delta_{\alpha\beta}}
{\f{1}{4}\epsilon^{\alpha\beta\gamma\delta}\Delta_{\alpha\beta}\Delta_{\gamma\delta}},
\ee
with $\f{1}{4}\eps\Delta\Delta=\f{1}{4}p^4\eps S S=\f{1}{2}p^4\eps s s$. We define the Dirac bracket as:
$$
\{f,g\}_D\equiv \{f,g\}- \{f,O_\mu\}(\Delta^{-1})^\mn\{O_\nu,g\}.
$$
Since $\{x_\mu, O_\nu\}=s_\mn$ and $\{p_\mu, O_\nu\}=0$, it is straightforward to compute\footnotemark:
$$
\{x_\mu,p_\nu\}_D=\eta_\mn,\quad
\{p_\mu,p_\nu\}_D=0,
$$
\be
\{x_\mu,x_\nu\}_D=\f{1}{2p^2}s_\mn + \phi_\mn(O_\alpha).
\ee
\footnotetext{The function of the constraint $O_\alpha$ appearing in the bracket $\{x,x\}$ is:
$$
\phi_\mn(O)=\f{4}{p^4(\eps s s)}\eps^{\alpha\beta\gamma\delta}s_{\mu\alpha}s_{\beta\nu}O_\gamma p_\delta.
$$}
We see that the spin of the particle induces a non-commutativity
of the particle position algebra at the classical level.

\subsection{About the quantification of the Dirac observables}

The Dirac observables $(L,X)$ form as in the spinless case  a $\so(4,1)$ algebra under the Poisson bracket. The quadratic Casimir
$\cc_2=\f12 LL-p^2XX$ can be computed exactly on-shell and we obtain $\cc_2=-T^2-\lambda^2$. The quartic Casimir $\cc_4=w_\mu
w^\mu$ is defined in term of the vector:
$$
w_\mu=\eps_{\mu \alpha\beta\gamma}X^\alpha(T)L^{\beta\gamma}.
$$

One can check that $\cc_4\ne 0$ unlike the spinless case. This
means that we are not restricted to simple representations anymore
as in the spinless case: we need to consider all functions
$L^2(\SO(1,4))$ and we can not restrict ourselves to the 5d light
cone. Thus, in the case of the spinning particle, we are
necessarily led to work with a 5d representation of the Dirac
observables.

\section*{Conclusion}

We have looked at  the relativistic particle from the perspective
of its algebra of (Dirac) observables. We have identified a set of Lorentz covariant
position observables, which turn out to be non-commutative. This non-commutativity
reflects the fact that one can not localize a massive quantum particle with a precision
better than its Compton length. We then showed that the particle admits in this context a
natural representation in five dimensions and could be quantized in terms of wave
functions on the 5d light cone. This allows a direct comparison with the free particle in
DSR (e.g.
\cite{GKKL}), which evolves in a non-commutative space-time and whose momentum lives in
the curved de Sitter space. Moreover, it turns out that the 5d light cone formulation is
subject to divergencies, which are naturally regulated by DSR. This allows a clear
understanding of how DSR arises as an extension of Special Relativity. In particular, it
lead us to an interpretation of the fifth coordinate as  proper time and of the fifth
moment as generating the Hamiltonian flow of the particle.

The case of the spinning particle deserves more attention. We have showed that the spin
induces an extra non-commutativity of the space-time coordinates. This complicates the
quantization of the algebra of observables and a full analysis of the 5d representation
of the spinning particle is still under investigation.

Finally, we also described a new representation of the Feynman diagram evaluations in
term of Dirac observables. We now hope to extend this approach to DSR and use these new
tools in order to build the scattering amplitudes and S-matrix of a Quantum Field Theory in DSR in a
consistent way.

\end{document}